\documentstyle[multicol,aps,epsfig]{revtex} 
\tighten       

\begin{document}
\draft 

\title{Chapman-Enskog method and synchronization
of globally coupled oscillators}
\author{ L.L. Bonilla\cite{bonilla:email}}
\address{Escuela Polit\'ecnica Superior, 
Universidad Carlos III de Madrid, Avda.\
Universidad 30, \\ 28911 Legan{\'e}s, Spain}
\date{July 10, 2000}

\maketitle
\begin{abstract}
The Chapman-Enskog method of kinetic theory is
applied to two problems of synchronization
of globally coupled phase oscillators. First, a
modified Kuramoto model is obtained in the limit
of small inertia from a more general model which
includes ``inertial'' effects. Second, a modified
Chapman-Enskog method is used  to derive the
amplitude equation for an $O(2)$ Takens-Bogdanov
bifurcation corresponding to the tricritical
point of the Kuramoto model with a bimodal
distribution of oscillator natural frequencies.
This latter calculation shows that the
Chapman-Enskog method is a convenient alternative
to normal form calculations. 
\end{abstract}


\begin{multicols}{2}
\narrowtext

\section{Introduction}

The Chapman-Enskog method (CEM) has long been used
to derive hydrodynamic equations of parabolic type
from kinetic equations of the Boltzmann type
\cite{chapman,cercignani}. The complexity of these
equations has perhaps hindered the realization
that the CEM is a powerful singular perturbation
method that can be used advantageously as a viable
alternative to multiple scales or normal form
calculations in different contexts. Motivated by
recent work on the subject of models of
synchronization of phase oscillators
\cite{bps,choi,ABS}, I shall derive here reduced
equations for two different Kuramoto models by
means of the CEM. 

The first example corresponds to the limit of
small inertia in a generalized Kuramoto model of
globally coupled phase oscillators
\cite{acebron2}:
\begin{eqnarray}
 \dot{\theta_{j}} &=&\omega_{j}\nonumber\\
m\,\dot{\omega_{j}} &=& -\omega_{j}+\Omega_{j}+
K\,r_{N} \sin(\psi_{N} -\theta_{j}) +
\xi_{j}(t),\label{1} \\ 
&&\quad \quad \quad \quad
\quad \quad \quad \quad \quad \quad \quad
j=1,\ldots,N,   \nonumber
\end{eqnarray}
where $\theta_{j}$, $\omega_{j}$ and $\Omega_{j}$
denote phase, frequency and natural frequency of
the $j$th oscillator, respectively. The natural
frequencies are distributed with probability
density $g(\Omega)$, which may have a single
maximum ({\em unimodal} distribution), or several
peaks ({\em multimodal} distribution). The
positive parameters $m$ and $K$ are the
``inertia'' and the coupling strength,
respectively. The complex order parameter defined
by
\begin{equation} r_{N}
e^{i\,\psi_{N}}=\frac{1}{N} \sum_{j=1}^{N} 
e^{i\,\theta_{j}},\label{2}
\end{equation}
measures phase synchronization: in the limit as
$N\to\infty$, $r_N>0$ if the oscillators are
synchronized and $r_N=0$ if not. Finally,
$\xi_{j}$'s are independent identically 
distributed Gaussian white noises, with $\langle
\xi_{j} \rangle =0,\,\langle\xi_{i}(t) \xi_{j}(s)
\rangle = 2D \delta_{ij} \delta(t-s)$. Without
white noise terms, these equations were proposed
to account for synchronization of biological
systems \cite{ermentrout,tanaka}. When the
inertial terms vanish, $m=0$, Eqs.\ (\ref{1}) and
(\ref{2}) are exactly the usual Kuramoto model
\cite{kuramoto,sakaguchi}. An important
technologically relevant application of this
model is the study of superconducting Josephson
junctions arrays \cite{wiesenfeld,park}. It has
been shown that the usual Kuramoto model (with
$m=0$) describes series arrays in the limiting
case of weak coupling and disorder of arrays with
zero junction capacitance \cite{wiesenfeld}. A
currently open problem is to derive a reduced
model for the averaged system in the limiting
case of weak coupling, weak disorder and nonzero
junction capacitance. In so far as capacitance
plays the same role as inertia in simple
electrical networks, the Kuramoto model with
inertia might be useful to understand the effect
of nonzero electrical capacitance in Josephson
junction arrays. 

In the limiting case of infinitely many
oscillators, models with mean-field coupling are
described by an evolution equation for the
one-oscillator probability density,
$\rho(\theta,\omega,\Omega,t)$, \cite{bon87}. For
the present model this equation is
\cite{acebron2}
\begin{eqnarray}
\frac{\partial \rho}{\partial t}&=&\frac{D}{m^2}
\frac{\partial^{2} \rho}{\partial
\omega^{2}} \nonumber\\
&-&\frac{1}{m}\frac{\partial}{\partial \omega}
[\big(-\omega+\Omega+K r
\sin(\psi-\theta)\big)\rho]
-\omega\frac{\partial \rho}{\partial\theta}\,,
\label{fpe}
\end{eqnarray}
where the order parameter is now given by
\begin{eqnarray}
r e^{i \psi}= \int_{0}^{2\pi}
\int_{-\infty}^{+\infty}
\int_{-\infty}^{+\infty}  e^{i \theta}
\rho(\theta,\omega,\Omega,t)
g(\Omega)\,d\Omega\, d\omega\, d\theta.
\label{order}
\end{eqnarray}
Equations (\ref{fpe}) and (\ref{order}) should be 
supplemented with appropriate initial and
boundary data ($\rho$ is $2\pi$-periodic in
$\theta$ and has suitable decay behavior as
$\omega\to\pm\infty$) plus the normalization
condition,
\begin{equation}
\int_0^{2\pi} \int_{-\infty}^{+\infty}
\rho(\theta,\omega, \Omega,t)\, d\omega\,
d\theta=1. 
\label{norma}
\end{equation}
An extense study of oscillator synchronization in
the model (\ref{fpe}) - (\ref{norma}) was carried
out in a previous paper \cite{ABS} by using
different analytical and numerical techniques.
The limiting case of small inertia, $m\to 0$, was
analyzed by an arbitrary three-mode truncation of
a mode-coupling expansion of $\rho$ \cite{ABS}.
It is rather intriguing that exactly the same
small inertia results had been obtained earlier by
Hong et al \cite{choi} using completely different
methods. Hong et al obtained first a reduced
Smoluchowski type equation, and then analyzed this reduced
equation. The derivation of the reduced equation
consisted of truncating a moment hierarchy for
the Fokker-Planck equation, as explained by
Schneider et al \cite{schneider}. Both procedures
yielded correctly the linear term of the amplitude
equation describing the synchronization
transition, but not the cubic term in this
equation. In this paper, we shall obtain for the
first time a consistent reduced Smoluchowski type
equation by using CEM. 

Our second example corresponds to the particular
case of setting $m=0$ and $g(\Omega) = {1\over
2}\, [\delta(\Omega-\Omega_0) + \delta(\Omega
+\Omega_0)]$ in (\ref{1}). Then the phase diagram
of the incoherent probability density
$\rho=1/(2\pi)$ contains a tricritical point
where a line of Hopf bifurcations intersects
tangentially a line of homoclinic orbits at zero
frequency. The resulting Takens-Bogdanov
bifurcation with $O(2)$ symmetry was analyzed in
\cite{bps} by multiple scales techniques,
\cite{kevorkian}. A shortcoming of these
techniques is that all terms in the corresponding
amplitude equation are necessarily of the same
order, while terms of different order are needed
to describe the Takens-Bogdanov point. We shall
show here how to adapt multiple scale ideas using
the CEM to overcome this difficulty. Although the
resulting amplitude equation is known, the
calculation presented here is of independent
interest to illustrate the possible applications
of our presentation of the CEM to deriving
amplitude and bifurcation equations containing
different asymptotic orders. 

The rest of the paper is as follows. The
Smoluchowski equation for the small inertia limit
of the Kuramoto model with inertia is derived in
Section II. Section III contains the application
of the CEM to finding bifurcation equations.
After explaining how the procedure works for the
well-known Hopf bifurcation, we describe how to
obtain the normal form corresponding to an
$O(2)$-symmetric Takens-Bogdanov bifurcation for
the bimodal Kuramoto model (without inertia). The
last Section contains our conclusions.

\section{Chapman-Enskog method for the
generalized Kuramoto model in the limiting case
of small inertia} 
To find a reduced equation, we should first
nondimensionalize (\ref{fpe}) and (\ref{order})
in an appropriate way. The main idea is that the
force terms and diffusion in velocity space
should be dominant. Then the probability density
$\rho$ rapidly reaches local equilibrium in
velocity and its slowly varying amplitude obeys a
Smoluchowski equation. This means that $\Omega$
and $K$ should have the same order as $\omega$,
and that the terms $m^{-1}(\omega\rho)_\omega$ and
$(D/m^2) \rho_{\omega\omega}$ should be of the
same order (subscripts mean partial derivative
with respect to the corresponding variable). If
we call $\omega_0$ a typical unit of velocity,
the latter balance yields the following velocity
scale
\begin{equation}
\omega_0 = \sqrt{{D\over m}}\,,\label{3}
\end{equation}
which is just the thermal velocity. The ratio of
$\omega\rho_\theta$ to $m^{-1}(\omega\rho)_\omega$
is of the order 
\begin{equation}
\epsilon = \sqrt{mD}\,,\label{epsilon}
\end{equation}
which will be our small dimensionless parameter.
Lastly, we shall choose the time unit so that
$\rho_t$ and $\omega\rho_\theta$ are of the same
order. This choice is dictated by the mechanics
of CEM (see below) and yields a time unit $t_0=
1/\omega_0$. Normalization condition and the 
definition of the order parameter dictate that
$g(\Omega)$ and $\rho$ are to be measured in
units of $1/\omega_0$ too. The angle $\theta$ is
already dimensionless. 

Recapitulating, $\omega$, $\Omega$ and $K$ are
measured in units of the thermal velocity
$\omega_0$, while $\rho$, $g(\Omega)$ and time
are measured in units of the reciprocal velocity
$1/\omega_0$. Then the dimensionless
Fokker-Planck equation is
\begin{eqnarray}
\frac{\partial}{\partial \omega}\left\{
\frac{\partial\rho}{\partial \omega} +
[\omega + K r \sin(\theta-\psi)-\Omega]\,
\rho\right\}\nonumber\\ 
=\epsilon\,\left(\frac{\partial\rho}{\partial t}
+ \omega\frac{\partial
\rho}{\partial\theta}\right) \,,
\label{-dfpe}
\end{eqnarray}
to be solved together with (\ref{order}),
(\ref{norma}), and appropriate periodicity, 
initial and decay conditions as $\omega\to\pm
\infty$. 

A different ``parabolic'' scaling is usual in
kinetic theory: in it the term $[K r\sin(\theta
-\psi)-\Omega]\rho_\omega$ is $O(\epsilon)$ and
the time derivative is $O(\epsilon^2)$. Using the
CEM with this scaling yields the standard
Kuramoto model with $m=0$ as a leading order
approximation, while higher order modifications
contain derivatives of order 4 in $\theta$ and
higher. We think the ``hyperbolic scaling''
(\ref{-dfpe}) will yield results which are valid 
over a much larger range of independent variables
and of parameters. 

\subsection{Chapman-Enskog method}
Setting $\epsilon=0$ in (\ref{-dfpe}), we find a
simple equation to be solved together with 
(\ref{order}) and (\ref{norma}). Its solution is a
displaced Maxwellian:
\begin{eqnarray}
\rho = {e^{-{V^{2}\over 2}}\over\sqrt{2\pi}}\,
P(\theta,\Omega,t),\label{4}\\
V = \omega + K r \sin(\theta-\psi)-\Omega,
\label{5}\\
r = \int_0^{2\pi}\int_{-\infty}^{\infty} e^{i(
\theta-\psi)} P(\theta,\Omega,t)\, d\theta\,
g(\Omega) d\Omega,\label{6}\\
\int_0^{2\pi} P(\theta,\Omega,t)\, d\theta = 1.
\label{22}
\end{eqnarray}
Notice that $P(\theta,\Omega,t)$ is an arbitrary
function of $\theta$ and $t$ except for
(\ref{22}). Furthermore, (\ref{4}) correspond to
a particular form of the initial conditions. The
Chapman-Enskog ansatz consists of assuming that
$\rho$ has the following asymptotic expansion 
\begin{eqnarray}
\rho = {e^{-{V^{2}\over 2}}\over\sqrt{2\pi}}\,
P(\theta,\Omega,t;\epsilon) + \sum_{n=1}^{\infty}
\epsilon^n \, \rho^{(n)}(\theta,\omega,\Omega;P).
\label{8}
\end{eqnarray}
Furthermore, we impose that the amplitude $P$
obeys an equation:
\begin{eqnarray}
{\partial P\over\partial t} = \sum_{n=0}^{\infty} 
\epsilon^n \, F^{(n)}(P),\label{9}
\end{eqnarray}
where $F^{(n)}$ are functionals of $P$ to be
determined as the procedure goes on. This
equation for $P$ is not explicitly written in the
usual presentations of CEM
\cite{chapman,cercignani}. Instead, the form of
this equation is guessed by writing equations for
the moments of $\rho$ and using gradient
expansions. We find this latter procedure more
confusing.

Insertion of (\ref{8}) and (\ref{9}) into the
equations and auxiliary conditions yields a
hierarchy of linear equations for the
$\rho^{(n)}$. Notice that the latter depend on
time only through their dependence on
$P$. The functionals $F^{(n)}(P)$ are determined
so that each equation (and set of auxiliary
conditions) for $\rho^{(n)}$ has a solution which
is bounded for all values of $\omega$, even as
$\omega\to\pm\infty$. Once a sufficient number of
$F^{(n)}$ is determined, (\ref{9}) {\em is the
sought amplitude equation}. Please notice that,
unlike results from the method of multiple
scales, terms in (\ref{9}) may be of different
order. 

Let us illustrate how the procedure works
by finding $F^{(0)}$ and $F^{(1)}$. Insertion of
(\ref{8}) and (\ref{9}) in (\ref{-dfpe}),
(\ref{order}) and (\ref{norma}) yields the
following hierarchy of linear equations: 
\begin{eqnarray}
{\cal L}\rho^{(1)} &=& {\omega e^{-{V^{2}\over
2}}\over\sqrt{2\pi}}\, \left[ P_\theta - VP
Kr^{(0)}\cos (\theta-\psi)\right]\nonumber\\
&+& {e^{-{V^{2}\over 2}}\over\sqrt{2\pi}}\,
\left[  F^{(0)} - VP K [\dot{r}^{(0)}\sin(\theta
-\psi)\right. \nonumber\\
&-& \left. r^{(0)}\dot{\psi}\cos(\theta-\psi)]
\right]\,,\label{10}\\
{\cal L}\rho^{(2)} &=& {e^{-{V^{2}\over 2}}\over
\sqrt{2\pi}}\, F^{(1)} + K r^{(1)}
\sin(\theta-\psi)\, \rho^{(1)}_\omega \nonumber\\
&+& \omega \rho^{(1)}_\theta + \rho^{(1)}_t
\big|_{P_{t}=F^{(0)}} \,,\label{11}
\end{eqnarray}
and so on. These equations are to be supplemented
by the normalization conditions (\ref{22}) and
\begin{eqnarray}
\int_0^{2\pi}\int_{-\infty}^{\infty}
\rho^{(n)} d\theta d\omega = 0, \label{12}
\end{eqnarray}
and the definitions:
\begin{eqnarray}
r^{(n)} = \int_0^{2\pi} \int_{-\infty}^{\infty}
\int_{-\infty}^{\infty} e^{i(\theta-\psi)}
\rho^{(n)} d\theta d\omega\, g(\Omega) d\Omega,
\label{13}\\
{\cal L}\rho^{(n)} = \left[V\rho^{(n)} +
\rho^{(n)}_\omega \right]_\omega - {V
e^{-{V^{2}\over 2}}\over\sqrt{2\pi}}\, K r^{(n)}
\sin(\theta-\psi), \label{14}
\end{eqnarray}
for $n=1,2,\ldots$. $V$ is now given by (\ref{5})
with $r=r^{(0)}$ given by (\ref{6}). Lastly,
$\dot{r}^{(0)}$ and $\dot{\psi}$ are calculated
by taking the time derivative of (\ref{6}) and
using the first term in (\ref{9}) to replace $P_t$
by $F^{(0)}$. 

Let us now consider (\ref{10}). Since the
$\omega$ integral of its left hand side is zero,
this equation has a solution only if the $\omega$
integral of its right hand side is zero. The
corresponding integrals are simplified by using
the symmetry of the Maxwellian and shifting
integration variables from $\omega$ to $V$. The 
vanishing of the integral of the right side yields
\begin{eqnarray}
F^{(0)} =  \left\{ [ K r^{(0)} \sin(\theta-\psi) -
\Omega]\, P\right\}_\theta . \label{15}
\end{eqnarray}
Notice that we needed $F^{(0)}$ in the right side
of (\ref{10}) for this equation to have an
appropriate solution. In turn, $F^{(0)}$
appeared in this equation due to our choice of
the time unit. Thus the time unit is dictated by 
the solvability conditions of the hierarchy of
equations generated by the CEM.

Inserting now (\ref{15}) in Eq.\ (\ref{10}), we
realize that the right hand side thereof is the
partial derivative of some expression with
respect to $\omega$. This greatly helps finding
the solution:
\begin{eqnarray}
\rho^{(1)} &=& {e^{-{V^{2}\over
2}}\over\sqrt{2\pi}}\, \left\{ {V^{2}-1\over 2}\,
PK r^{(0)} \cos(\theta-\psi) \right.\nonumber\\
&+&
V\left[ KP\left( \dot{r}^{(0)}\sin(\theta-\psi)
- r^{(0)}\dot{\psi}\cos(\theta-\psi)
\right.\right. \nonumber\\ 
&-& \left.\left.\left. [K r^{(0)}
\sin(\theta-\psi) -\Omega]\,  r^{(0)}
\cos(\theta-\psi)\right) - P_{\theta}
\right]\right\} ,    \label{16}
\end{eqnarray}
which satisfies (\ref{12}) and yields $r^{(1)} =
0$. Notice that a term of the form (\ref{4})
(satisfying $\int_0^{2\pi} P\, d\theta = 0$)
could have been added to the solution (\ref{16}).
However, all such terms are already contained in
the ansatz (\ref{8}) with $P =
P(\theta,\Omega,t;\epsilon)$, and we shall
therefore omit them. 

To find $F^{(1)}$, we insert (\ref{16}) in
(\ref{11}) and use the solvability condition for
this equation. Simplifications arise from the
identities 
$$\int_{-\infty}^{\infty} \rho^{(1)}_t d\omega = 
{\partial\over\partial t} \int_{-\infty}^{\infty}
\rho^{(1)} d\omega = 0,
$$
and  
$$\int_{-\infty}^{\infty} \omega \rho^{(1)}_\theta
d\omega =  {\partial\over\partial \theta}
\int_{-\infty}^{\infty} \omega
\rho^{(1)}\, d\omega.
$$
The result is 
\begin{eqnarray}
F^{(1)} = {\partial\over\partial\theta} \left\{
P_{\theta} + \left( [K r^{(0)}\sin(\theta-\psi)
-\Omega + \dot{\psi}]\right. \right.\nonumber\\
\times\left.\left. K r^{(0)}
\cos(\theta-\psi) - K\dot{r}^{(0)}
\sin(\theta-\psi) \right)\, P \right\} \,.
\label{17}
\end{eqnarray}
In this expression, $\dot{r}^{(0)}$ and
$\dot{\psi}$ can be found from their definition
and Eq.\ (\ref{15}). Integration by parts yields: 
\begin{eqnarray}
\dot{r}^{(0)} + i r^{(0)}\dot{\psi} = {K
r^{(0)}\over 2}\,
\left(1-
\langle  e^{2i(\theta-\psi)} \rangle\right) + i
\langle \Omega e^{i(\theta-\psi)} \rangle
\nonumber\\
\Longrightarrow 
 \dot{r}^{(0)} = K r^{(0)} \langle\sin^2 (\theta-
\psi)\rangle - \langle\Omega\sin(\theta-
\psi)\rangle,\nonumber\\  
r^{(0)}\dot{\psi} = \langle \Omega
\cos(\theta-\psi)\rangle - {K r^{(0)}\over 2}\,
\langle\sin[2 (\theta-\psi)]\rangle,\label{18}
\end{eqnarray}
where
\begin{eqnarray}
\langle f(\theta,\Omega)\rangle \equiv
\int_0^{2\pi} \int_{-\infty}^{\infty}
f(\theta,\Omega)\, P(\theta,\Omega,t;\epsilon)\,
d\theta\, g(\Omega)\, d\Omega\,. \label{19}
\end{eqnarray}
We can now insert (\ref{15}), (\ref{17}) and
(\ref{18}) into (\ref{9}) to obtain the sought
Smoluchowski equation for $P$:
\begin{eqnarray}
P_t &-& {\partial\over\partial\theta} \left\{
\left( [Kr^{(0)}\sin(\theta-\psi) -\Omega]\, [1+
\epsilon K r^{(0)}\cos(\theta-\psi)] 
\right.\right. \nonumber\\
&-& \epsilon 
\left({K r^{(0)}\over 2}\, \langle\sin[2
(\theta-\psi)]\rangle-\langle\Omega
\cos(\theta-\psi)\rangle \right)\nonumber\\
&\times&  K \cos(\theta-\psi) -   
\epsilon K\sin(\theta-\psi)\nonumber\\
&\times& \left.\left[Kr^{(0)}\langle\sin^2
(\theta-\psi)\rangle - \langle\Omega\sin(\theta-
\psi)\rangle  \right]\right)\, P \nonumber\\
&+& \left. \epsilon P_{\theta}\right\} = 0\,.
\label{20}
\end{eqnarray}

Restoring the original dimensional units, this
equation becomes:
\begin{eqnarray}
P_t &-& {\partial\over\partial\theta} \left\{
\left( [Kr\sin(\theta-\psi) -\Omega]\, [1+
m K r\cos(\theta-\psi)] 
\right.\right. \nonumber\\
&-& m K \cos(\theta-\psi)
\left({K r\over 2}\, \langle\sin[2
(\theta-\psi)]\rangle\right. \nonumber\\
&-&\left. \langle\Omega
\cos(\theta-\psi)\rangle \right) - m K
\sin(\theta-\psi)\nonumber\\
&\times& \left. \left[Kr\langle\sin^2
(\theta-\psi)\rangle - \langle\Omega\sin(\theta-
\psi)\rangle  \right]\, \right)\, P \nonumber\\
&+&\left. D P_{\theta}\right\} = 0\,.  \label{21}
\end{eqnarray}
to be solved together with (\ref{6}), definition
(\ref{19}), normalization condition (\ref{22}),
$2\pi$-periodicity in $\theta$ and appropriate
initial conditions. 

Equation (\ref{21}) is the main result of this
Section. If $m=0$, the usual Kuramoto model for
phase oscillators is recovered. Comparing this
result with Hong et al's Equation (5), we see that
the second and third terms proportional to $m$ in
the drift current were missing and that their
equation contained an additional diffusive term.
The latter would be order $\epsilon^2$ in our
scheme. These differences can be tracked to the
fact that Hong et al followed a previous paper
\cite{schneider}, whose authors used a moment
method with arbitrary closure assumptions to
derive the Smoluchowski equation. 

\section{Derivation of bifurcation equations by
the Chapman-Enskog method}
First of all, we shall illustrate the application
of the CEM to a well-known case: deriving the
amplitude equation for the usual Hopf
bifurcation. Next the much more complicated
calculation of the amplitude equation for the
tricritical point of the bimodal Kuramoto model
will be tackled.

\subsection{The usual Hopf bifurcation}
Let us consider a system of $n$ ordinary
differential equations for a $n$-component vector
unknown $u(t;\alpha)$:
\begin{eqnarray}
{du\over dt} = f(u;\alpha); \label{ho1}
\end{eqnarray}
$\alpha$ is the bifurcation parameter. we shall
assume that $u=0$ is a linearly stable solution if
$\alpha<0$ and that it is unstable if $\alpha>0$.
All eigenvalues of $L(\alpha)\equiv \partial
f(0,\alpha)/\partial u$ except $\lambda(\alpha)$
and its complex conjugate have negative real
parts in a neighborhood of $\alpha=0$. We have
$\lambda(\alpha)\sim i\omega_0 + \alpha 
\lambda'(0)$ as $\alpha\to 0$, with
Re$\lambda'(0)>0$. The linearized equation of
(\ref{ho1}) about $u=0$, $\alpha=0$ has the
solution $u= A\,\phi_0\,  e^{i\omega_{0}t}+$cc,
plus terms which decay exponentially fast to zero
as time increases. $\phi_0$ is the eigenvector 
corresponding to $i\omega_0$ and $A$ is a complex
constant.

To apply CEM to this problem, we shall assume
\begin{eqnarray}
\alpha = \alpha_1 \varepsilon + \alpha_2
\varepsilon^2  + O(\varepsilon^3),\nonumber\\  
u = \varepsilon\, A(t;\varepsilon)\,
\phi_0\, e^{i\omega_{0}t} + \mbox{cc} +
\sum_{n=2}^{\infty}\varepsilon^n\, 
u^{(n)}(t;A,\overline{A}), \label{ho2}
\end{eqnarray}
where the amplitude $A$ obeys the ansatz 
\begin{eqnarray}
{dA\over dt} = \sum_{n=1}^{\infty}\varepsilon^n\, 
F^{(n)}(A,\overline{A}).\label{ho3}
\end{eqnarray}
Here $\overline{A}$ is the complex conjugate of
$A$ and $cc$ means complex conjugate of the
preceding term. The parameter $\varepsilon$
measures the size of the bifurcating amplitude
and its relation to the bifurcation parameter
$\alpha$ is determined by finding the
coefficients $\alpha_j$. Notice that $A$ varies
slowly with time (on a scale $\epsilon^2 t$, as we
will see). The functions $u^{(n)}$ depend on a
fast scale $t$ corresponding to stable
exponentially decaying modes, and {\em on a slow
time scale through their dependence on} $A$. All
terms in (\ref{ho2}) which decrease exponentially
in time will be omitted. $\alpha_n$, $F^{(n)}$ are
determined so that the solutions $u^{(n)}$ are
bounded as $t\to\infty$ (on the fast time scale),
for fixed $A$. Substitution of (\ref{ho2}) and
(\ref{ho3}) in (\ref{ho1}) yields the following 
hierarchy of linear equations:
$$[L(0) - i\omega_0]\, \phi_0 =0,$$
\begin{eqnarray}
{du^{(2)}\over dt} - L(0) u^{(2)}= 
[\alpha_1 A\, L'(0)- F^{(1)}]\, \phi_0
e^{i\omega_{0}t} + \mbox{cc} \nonumber\\  
+ {1\over 2} f_{uu}(0;0): (A\phi_0
e^{i\omega_{0}t} + \mbox{cc})^2 ,
\label{ho4}
\end{eqnarray}
\begin{eqnarray}
{du^{(3)}\over dt} - L(0) u^{(3)}= \alpha_1 
(\ldots) + [\alpha_2 A\, L'(0)-
F^{(2)}]\nonumber\\ 
\times \phi_0 e^{i\omega_{0}t} 
+ \mbox{cc} + {1\over 2} f_{uu}(0;0):
u^{(2)}(A\phi_0  e^{i\omega_{0}t} + \mbox{cc})
\nonumber\\
+ {1\over 6} f_{uuu}(0;0): (A\phi_0 
e^{i\omega_{0}t} + \mbox{cc})^3 ,
\label{ho5}
\end{eqnarray}
and so on. Here $f_{uu}(0;0)$ is a $n\times n$
matrix, colon sign means tensor contraction, etc.
The first equation holds identically due to the
definitions of $\phi_0$ and $i\omega_0$. The
other equations should be solved for bounded 
$u^{(n)}$ as $t\to\infty$. Their solutions should
not contain terms of the form $B\phi_0
e^{i\omega_{0}t} + $cc, solving the corresponding
linear homogeneous problem. The reason for this
latter requirement is that all such terms are
already contained in $A(t;\varepsilon)$. 

Eq.\ (\ref{ho4}) yields 
$$\alpha_1=0, \quad F^{(1)}=0,$$
\begin{eqnarray}
u^{(2)}= {A^{2}\over 2}\, e^{i 2\omega_{0}t} [2i
\omega_0 I - L(0)]^{-1} f_{uu}(0;0):\phi_0^2
\nonumber\\
+ \mbox{cc} - |A|^2 L(0)^{-1} f_{uu}(0;0): 
\phi_0 \overline{\phi_{0}}\,. \label{ho6}
\end{eqnarray}
($I$ is the identity matrix of order $n$).
Solvability of (\ref{ho5}) yields 
\begin{eqnarray}
F^{(2)} = \alpha_2 \lambda'(0) A - \mu\, A |A|^2,
\label{ho7}
\end{eqnarray}
where 
\begin{eqnarray}
\lambda'(0) = {\langle \phi_{0}^{\dag},
L'(0)\phi_{0}\rangle\over
\langle \phi_{0}^{\dag},\phi_{0}\rangle},
\label{ho8}\\
\mu = \langle
\phi_{0}^{\dag},\phi_{0}\rangle^{-1}\,
\left\{- {1\over 2}\, \langle \phi_{0}^{\dag},
f_{uuu}(0;0): \phi_0^2 \overline{\phi_{0}} 
\rangle\right.\nonumber\\
+ \langle
\phi_{0}^{\dag},f_{uu}(0;0):\phi_{0}  L(0)^{-1}
f_{uu}(0;0):\phi_0 \overline{\phi_{0}}
\rangle\nonumber\\
\left. + {1\over 2}\, \langle
\phi_{0}^{\dag}, f_{uu}(0;0): \overline{\phi_{0}}
[L(0)-i2\omega_0 I]^{-1} f_{uu}(0;0): \phi_0^2
\rangle\right\} .\label{ho9} 
\end{eqnarray}
Here $\phi_{0}^{\dag}$ is the eigenvector
corresponding to the adjoint problem
$$L^{\dag}(0)\phi_{0}^{\dag} = i\omega_0 \phi_{0}^{\dag}.$$
Substituting these results in (\ref{ho3}), the
following amplitude equation is found:
\begin{eqnarray}
{dA\over dt} = \varepsilon^2\, \alpha_2 \lambda'(0)\,
A - \varepsilon^2\, \mu\, A\, |A|^2 +
O(\varepsilon^{3}).  \label{ho10}
\end{eqnarray}

These formulas for the bifurcation equation and
its coefficients have been obtained many times
before; see \cite{poore} for an equivalent
explicit determination of $\mu$, whose real part
decides whether the bifurcation is sub or
supercritical. If Re$\mu =0$, we should calculate
higher order terms in the Chapman-Enskog expansion
(\ref{ho3}). The systematic way in which the CEM
method yields such terms is a great advantage
with respect to other methods such as multiple
scales \cite{kevorkian}. 

Similar ideas can be used to derive amplitude
equations for pattern forming systems governed by
partial differential equations \cite{cross}. In
such cases, we have to rescale appropriately
space variables and assume that the $F^{(n)}$ also
depend on spatial derivatives of $A$. 

\subsection{Symmetric Takens-Bogdanov
bifurcation} 
Our starting point in this Subsection is the
standard Kuramoto model with bimodal natural  
frequency distribution, i.e., (\ref{21}) with
$m=0$ and $g(\Omega) = {1\over 2}[\delta
(\Omega-\Omega_0) + \delta(\Omega+\Omega_0)]$.
The phase diagram of the incoherent solution $P=
1/(2\pi)$ was depicted in Figure 1 of \cite{bps}.
At the tricritical point, $P = (K/D=4,\Omega_{0}
/D = 1)$ a branch of Hopf bifurcations coalesce
with a branch with stationary bifurcations and a
branch of homoclinic orbits, in a $O(2)$-symmetric
Takens-Bogdanov bifurcation point. A method of
multiple scales was used in \cite{bps} to analyze
this complicated bifurcation. This method was not
completely satisfactory because it led to a
couple of amplitude equations whose solutions
were later interpreted as the two terms of a
normal form expansion \cite{bps}. Let us show how
a modified CEM leads directly to the amplitude
equation. Near the tricritical point, we may
define an expansion parameter $\varepsilon$ so
that
\begin{eqnarray}
K = K_{c} + K_{2} \varepsilon^2 +
O(\varepsilon^3),  \ \ \
 \Omega_{0} = \Omega_{0c} + \Omega_{2}
\varepsilon^2 + O(\varepsilon^3) \nonumber\\    
 (K_{c} = 4 D, \  \Omega_{0c} = D). 
\label{23} 
\end{eqnarray}
The basic slow time scale near the tricritical
point is $T=\varepsilon t$, and the method of
multiple scales show that resonant terms appear
in the equations of order $\varepsilon^{3}$ and
higher \cite{bps}. Borrowing from the results of
that reference (which includes making an
exponential ansatz for $P$), we shall make the
following Chapman-Enskog ansatz:
\begin{eqnarray}
P(\theta,\Omega,t;\varepsilon) = \frac{1}{2 \pi}
\exp \left\{\varepsilon\, \frac{
A(T;\varepsilon)}{D + i\Omega} e^{i\theta} + cc
\right. \nonumber\\ 
\left. + \sum_{j = 2}^{4} \varepsilon^j
\sigma_{j}(\theta,t,T;A) + O(\varepsilon^5)
\right\}\,, \label{24}\\
A_{TT} = F^{(0)}(A) + \varepsilon F^{(1)}(A) +
O(\varepsilon^2). \label{25}
\end{eqnarray}
Terms which decay exponentially on the fast time
scale, $t$, will be systematically omitted. The
equation for $A$ is second order [not first
order as (\ref{ho3})] because resonant terms
appear at $O(\varepsilon^3)$ for the first time.
Inserting these equations in  (\ref{21}) (with
$m=0$), we obtain the following hierarchy of
linear equations:  
\begin{eqnarray}
{\cal L} \left( \sigma_{2} +
\frac{\sigma_{1}^2}{2} \right) = -4 D
\partial_{\theta} \left\{ \sigma_{1} \mbox{Im} \ 
e^{-i \theta} \langle e^{i \theta'},\sigma_{1}
\rangle \right\} \nonumber\\ - {A_{T}
e^{i\theta}\over D + i\Omega} + cc, \nonumber\\
\label{29}\\
\int_{0}^{2 \pi} \left( \sigma_{2} +
\frac{\sigma_{1}^2}{2}\right)  d\theta = 0,
\nonumber
\end{eqnarray}

\begin{eqnarray}
{\cal L} \left( \sigma_{3} + \sigma_{1}
\sigma_{2} + \frac{\sigma_{1}^{2}}{6} \right)
= -4 D \,\partial_{\theta} \left\{ \sigma_{1}
\mbox{Im} \ e^{-i\theta} \langle
e^{i\theta'}, \right.
\nonumber\\  
\sigma_{2} + \frac{\sigma_{1}^{2}}{2}
\rangle + \left(\sigma_{2} +
\frac{\sigma_{1}^{2}}{2}
\right)\mbox{Im}\, e^{-i\theta} \langle 
e^{i\theta'},\sigma_{1} \rangle \nonumber\\
\left. + \Omega_{2}\, \mbox{Im} \ e^{-i\theta}
\langle e^{i \theta'},\sigma_{1}\rangle' 
\right\} \nonumber\\
- K_{2} \partial_{\theta}
\mbox{Im}\ e^{-i\theta} \langle e^{-i\theta'},
\sigma_{1}\rangle
 - \partial_{T} \left(
\sigma_{2} + \frac{\sigma_{1}^{2}}{2} \right) , 
\nonumber\\
 \label{29bis}\\
\int_{0}^{2 \pi} \left( \sigma_{3} + \sigma_{1}
\sigma_{2} +
	   \frac{\sigma_{1}^{2}}{6} \right) d\theta = 0,
\nonumber
\end{eqnarray}

\begin{eqnarray}
{\cal L} \left( \sigma_{4} + \sigma_{1}
\sigma_{3} + \frac{\sigma_{2}^2}{2} +
\frac{\sigma_{1}^2 \sigma_{2}}{2} +
\frac{\sigma_{1}^4}{4!} \right) = \nonumber\\
-4 D \partial_{\theta} \left\{ \sigma_{1}
\mbox{Im} \ e^{-i \theta}\langle e^{-i \theta'},
\sigma_{3} + \sigma_{1}\sigma_{2}	+
\frac{\sigma_{1}^3}{6} \rangle \right.
\nonumber\\   
 + \left( \sigma_{3} + \sigma_{1}
\sigma_{2} + \frac{\sigma_{1}^3}{6} \right)
\mbox{Im} \ e^{-i \theta} \langle e^{i\theta'},
\sigma_{1} \rangle \nonumber\\
+ \Omega_{2}\, [\mbox{Im} \ e^{-i\theta}\langle
e^{i \theta'}, \sigma_{2} +
\frac{\sigma_{1}^2}{2} \rangle'	 
+ \sigma_{1} \mbox{Im} \
e^{-i \theta} \langle	e^{i \theta'}, \sigma_{1}
\rangle'] \nonumber\\
\left. + \left( \sigma_{2} + 
\frac{\sigma_{1}^2}{2} \right) \mbox{Im} \ e^{-i
\theta} \langle e^{i\theta'}, \sigma_{2} +
\frac{\sigma_{1}^2}{2} \rangle  \right\}
\nonumber\\
- K_{2}\, \partial_{\theta} \left\{ \mbox{Im}\
e^{-i \theta} \langle e^{i \theta'},\sigma_{2} +
\frac{\sigma_{1}^2}{2} \rangle\right.\nonumber\\
\left. + \sigma_{1}
\mbox{Im} \ e^{-i\theta} \langle e^{i \theta'},
\sigma_{1} \rangle \right\}
\nonumber\\
- \partial_{T}\left( \sigma_{3} + \sigma_{1}
\sigma_{2} + \frac{\sigma_{1}^2}{2} \right),
\nonumber\\
\label{30}\\
\int_{0}^{2 \pi} \left( \sigma_{4} + \sigma_{1}
\sigma_{3} + \frac{\sigma_{2}^2}{2} +
\frac{\sigma_{1}^2 \sigma_{2}}{2} +
\frac{\sigma_{1}^4}{4!} \right) d \theta = 0.
\nonumber
\end{eqnarray}
Here
\begin{eqnarray}
{\cal L} \sigma_{n} = (\partial_{t}-D
\partial_{\theta}^2 + \Omega 
\partial_{\theta})	\sigma_{n}\nonumber\\
+ 4 D \partial_{\theta} \left\{ \mbox{Im}\
e^{-i \theta} \langle e^{i
\theta'}, \sigma_{n} \rangle \right\}\,,
\label{28}
\end{eqnarray}
and we have used the abbreviations $\sigma_1 =
{A(T;\varepsilon)\, e^{i\theta}\over D + i\Omega}
+ cc$ and 
\begin{eqnarray}
\langle \alpha(\theta,\Omega),\beta(\theta,
\Omega) \rangle	= \frac{1}{2 \pi} \int_{0}^{2
\pi} \int_{- \infty}^{+ \infty} \alpha(\theta,
\Omega) \beta(\theta, \Omega)\nonumber\\
 g(\Omega)\, d\theta\, d \Omega,\label{26}\\
\langle \alpha(\theta, \Omega), \beta(\theta,
\Omega) \rangle'	= \frac{1}{2 \pi} \int_{0}^{2
\pi} \int_{- \infty}^{+ \infty} \alpha(\theta,
\Omega) \beta(\theta, \Omega)\nonumber\\
g_{\Omega_{0}}'(\Omega)\, d\theta\, d\Omega,
\label{32}
\end{eqnarray}
where
\begin{eqnarray}
g_{\Omega_{0}}' (\Omega) = \frac{1}{2} \left[
\delta'(\Omega + \Omega_{0}) - \delta'(\Omega -
\Omega_{0}) \right],\label{33}\\
( \Omega_{0} = \Omega_{0c} = D).\nonumber
\end{eqnarray}
The ansatz (\ref{25}) has not yet been
inserted in (\ref{29}) and (\ref{30}) in order
not to complicate further these equations. We
shall keep in mind that these equations will have
to be modified later when the solution of
(\ref{29}) is inserted in (\ref{29bis}) and
(\ref{30}). The linear equations in the preceding
hierarchy should have solutions periodic in
$\theta$. A solution of the homogeneous equation 
${\cal L}U=0$ could be added to the solutions of
the linear nonhomogeneous equations in the
preceding hierarchy. However, all such terms are
to be omitted, for the amplitude
$A(T;\varepsilon)$ already takes care of them. 

The solution of (\ref{29}), 
\begin{equation}
{\cal L} \left( \sigma_{2} +
\frac{\sigma_{1}^2}{2} \right) = \left[-
\frac{A_{T} e^{i \theta}}{D + i \Omega} +
\frac{2 A^{2}e^{2 i \theta} }{D + i
\Omega}\right]  + cc,\label{35}
\end{equation}
is
\begin{equation}
\sigma_{2} + \frac{\sigma_{1}^2}{2} = \left[
\frac{A^{2} e^{2 i\theta} }{(D + i \Omega)(2 D + 
i\Omega)} - \frac{A_{T} e^{i \theta} }{(D + i
\Omega)^2} \right] + cc .\label{36}
\end{equation}
Note that one term proportional to $e^{i\theta}/
(D+i\Omega)$ and other terms decaying on
the fast scale $t$ are solutions of ${\cal L}U=0$
and could have been added to (\ref{36}). According
to what was said above, all such terms are to be
omitted. 

We now insert (\ref{36}) in (\ref{29bis}) and
(\ref{30}) together with the CE ansatz
(\ref{25}). The solution of (\ref{29bis}) is 
\begin{eqnarray}
\sigma_{3} + \sigma_{1} \sigma_{2} +
\frac{\sigma_{1}^3}{6}	 = \left[ \frac{K_{2} - 4
\Omega_{2}}{4 D (D + i \Omega)} A +
\frac{F^{(0)}}{(D + i \Omega)^3}\right.
 \nonumber\\
\left. - \frac{A |A|^{2}}{(D + i \Omega)^{2}
(2 D + i \Omega)} \right]\, e^{i \theta}  + cc 
\nonumber\\
-\frac{ A A_{T} \left(\frac{1}{D+i\Omega}
+ \frac{1}{2 D+ i \Omega} \right)}{(D + i
\Omega)(2 D + i\Omega)} e^{2 i \theta}  + cc 
\nonumber\\
+ \frac{A^{3} e^{3 i \theta}}{(D + i\Omega)(2
D + i \Omega)(3 D + i \Omega)} + cc , \label{37}
\end{eqnarray}
From this we obtain $\sigma_{3}$, and finally,
from (\ref{30}), $\sigma_{4}$.  To obtain the
leading order approximation, we only need to
determine $A(T;\varepsilon)$. Now, (\ref{37})
holds provided that the {\it nonresonance
condition}  (needed to remove secular terms)
\begin{equation}
\langle \frac{1}{D + i \Omega}, P(\Omega,T;A)
\rangle = 0 \label{38}
\end{equation}
holds, where $P(\Omega, T;A)$ denotes
the  coefficient of $e^{i \theta}$ on the
right-hand side of (\ref{29bis}) \cite{bps}.
Equation (\ref{38}) yields
\begin{equation}
F^{(0)} = \frac{D}{2} (K_{2} - 4 \Omega_{2}) A +
\frac{2}{5} |A|^2 A = 0.\label{39}
\end{equation}

The function $F^{(1)}$ is determined from a
similar nonresonance condition for (\ref{30}):
the coefficient $Q(\Omega, T;A)$ of $e^{i \theta}$
on the right-hand side of (\ref{30}) should also
satisfy (\ref{38}). The result is 
\begin{eqnarray}
F^{(1)} = \frac{K_{2}}{2} A_{T} -
\frac{\left(|A|^2 A \right)_{T}}{5 D} -
\frac{23}{25 D} |A|^2 A_{T}. \label{40}
\end{eqnarray}
Insertion of Equations (\ref{39}) and (\ref{40})
into (\ref{25}) yields the sought amplitude
equation: 
\begin{eqnarray}
A_{TT} - \frac{D}{2} (K_{2} - 4 \Omega_{2}) A -
\frac{2}{5} |A|^2 A	 =\nonumber\\
\varepsilon\,\left(\frac{K_{2}}{2}  A_{T} -
\frac{23}{25 D} |A|^2 A_{T} - \frac{1}{5 D} (|A|^2
A)_{T}\right) + O(\varepsilon^2).
\label{44}
\end{eqnarray}
This equation is the  scaled ``normal form''
studied by Dangelmayr and Knobloch in
\cite{DK} [cf.\ their equations (3.3),  p.\
2480]. The general analysis developed in that
Reference for general scaled normal forms was
employed in \cite{bps} to study (\ref{44}) and
will not be repeated here. 

\section{Conclusions}
We have applied a modified Chapman-Enskog method
to two problems related to Kuramoto models of
synchronization of globally coupled phase
oscillators. First of all, we found a consistent
two-term Smoluchowski approximate equation for a
model of oscillators with inertia in the limit of
small inertia (as $mD\to 0+$). Second, we
modified the Chapman-Enskog method to find
directly the scaled normal form corresponding to
the $O(2)$-symmetric Takens-Bogdanov bifurcation
at the tricritical point of a standard Kuramoto
model with bimodal distribution of oscillator
natural frequencies. Key ingredients of the CEM
are: (i) solving a zeroth-order problem whose
solution is determined up to certain amplitude
functions; (ii) assuming an expansion for the
solution all whose higher order terms depend on
(slow) time through the amplitude functions only;
(iii) assuming that the right sides of the
equations of motion for the amplitude functions
are expansions whose coefficients are functionals
of the amplitudes. These coefficients are
determined by appropriate solvability conditions
for the hierarchy of linear equations resulting
from insertion of all these assumptions in the
original equations. Collecting the desired number
of coefficients, we obtain approximate equations
of motion for the amplitude functions as the
result of the method. I believe that the
techniques explained in the present paper will be
useful in many other problems of physical
interest. 

\acknowledgements
I thank John C.\ Neu for an original presentation
on deriving the classical Smoluchowski equation,
Juan Soler for calling my attention to hyperbolic
limits of kinetic equations and Stephen W.\
Teitsworth for comments on the manuscript. I also
thank Prof.\ M.Y.\ Choi for bringing Ref.\
\cite{choi} to my attention. The present work was
financed through the Spanish DGES grant
PB98-0142-C04-01 and the European Union TMR
contract ERB FMBX-CT97-0157.

\end{multicols}
\end{document}